\documentclass[showpacs,amsmath,amssymb,prd,superscriptaddress,floatfix,twocolumn]{revtex4-1}
\usepackage{graphicx}
\usepackage{dcolumn}
\usepackage{bm}

\begin{document}

\title{A multi-scale perturbative approach to  $SU(2)$-Higgs classical dynamics: \\
stability of nonlinear plane waves and bounds of the Higgs field mass}
\author{V. Achilleos}
\author{F.K. Diakonos}
\author{D.J. Frantzeskakis}
\author{G.C. Katsimiga}
\author{X.N. Maintas}
\author{C.E. Tsagkarakis}
\affiliation{Department of Physics, University of Athens, GR-15874 Athens, Greece}
\author{A. Tsapalis}
\affiliation{Department of Physics, University of Athens, GR-15874 Athens, Greece}
\affiliation{Hellenic Naval Academy, Hatzikyriakou Avenue, Pireaus 185 39, Greece}


\begin{abstract}

We study the classical dynamics of $SU(2)$-Higgs field theory using multiple scale perturbation theory. In the spontaneously broken phase, assuming small perturbations of the Higgs field around its vacuum expectation value, we derive a nonlinear Schr\"{o}dinger equation and study the stability of its nonlinear plane wave solutions. The latter, turn out to be stable only if the Higgs amplitude is an order of magnitude smaller than that of the gauge field. In this case, the Higgs field mass possesses some bounds which may be relevant to the search for the Higgs particle at ongoing experiments.
\end{abstract}
\maketitle

{\it Introduction.} One of the main goals of the current collider experiments (Large Hadron Collider (LHC) at CERN, Tevatron at Fermilab) is the detection of the Higgs particle \cite{Higgs1964}, the missing piece in the experimental verification of the Standard Model \cite{Djouadi2008}. Existing bounds on the properties of the Higgs particle originate exclusively from quantum corrections of the Standard Model or its extensions (see Ref.~\cite{Schuecker2010} and references therein). This implies a different treatment of the Higgs field as compared to the electroweak gauge fields since, for the latter, a rough estimation of their properties is obtained already at the classical (Born) level. The reason for this discrimination is the fact that the weak boson mass is determined only from the vacuum expectation value (vev) of the Higgs field, while the Higgs mass depends explicitly also on the unknown coupling constant determining its self-interaction. In fact, the self-interaction corresponds to the presence of nonlinear terms in the Higgs potential. Usually, the nonlinear terms in the $SU(2)$-Higgs Lagrangian are treated as perturbation of an underlying linear theory allowing, this way, the straightforward canonical quantization of the electroweak theory. From the classical point of view, however, this treatment is insufficient. The reason is that such a perturbation scheme usually develops instabilities due to the emergence of secular terms at higher orders (see, e.g., Ref.~\cite{Jeffrey1982}).

In the present work we study the stability of plane waves in the presence of the nonlinear terms in the $SU(2)$-Higgs model --
a problem that, to the best of our knowledge, has not been addressed so far in the literature.
There is strong motivation in performing such a study. Generally, classical solutions may describe sufficiently the effective dynamics of nonlinear quantum fields since, in this case,  energy can be transferred from high frequency modes to long-wavelength excitations. In particular, they are important for the description of condensates, as indicated by an analogous treatment for the dynamics of the QCD chiral condensate in terms of classical pion fields \cite{Anselm1991}.
Finally, stability is a prerequisite for using classical solutions as a basis for the quantization of a classical field theory.

In our approach the classical equations of the $SU(2)$-Higgs model are solved approximately, using multiple scale perturbation theory \cite{Jeffrey1982} to handle the nonlinear terms.
This technique employs slow temporal and spatial scales to reduce the original nonlinear system to a simpler one, thus allowing the understanding of gross features of the original problem and enabling the consistent treatment of the secular terms. We note in passing that such perturbative methods are commonly used in a variety of physical contexts, ranging from water waves to nonlinear optics, condensed-matter physics, etc. In our case, assuming small perturbations of the Higgs field around its vev, we derive a set of coupled nonlinear equations for the evolution of the Higgs and the gauge boson fields. If the Higgs field amplitude is much smaller compared to the gauge boson field, these equations decouple and obey a nonlinear Schr\"{o}dinger (NLS) equation. The latter, possesses nonlinear plane wave (among other, soliton-type) solutions, which have been already discussed in literature \cite{Coleman1977,Savvidy1979}. Here, we will focus on the stability properties of these solutions and show that the stability condition for the nonlinear plane waves leads to restrictions in the values of the ratio of the Higgs to the gauge boson field mass.

{\it Formulation and Setup.} In our treatment we neglect the mixing with the electromagnetic $U(1)$ gauge field since it does not couple directly to the Higgs field. As a consequence, the mass of the three components of the $SU(2)$ field is taken to be the same. We also do not consider here the fermionic sector. The $SU(2)$-Higgs field dynamics is described by the Lagrangian:
\begin{equation}
{\mathcal{L}}=-\frac{1}{4} F^a_{\mu \nu} F^{a, \mu \nu} + (D_\mu \phi)^{\dagger}(D^\mu \phi)
- V(\Phi^{\dagger} \Phi),
\label{eq:eq1}
\end{equation}
where $F^a_{\mu \nu}$ is the SU(2) field strength tensor, $D_\mu=\partial_\mu +igA^a_\mu {\sigma^a \over 2}$ is the associated covariant derivative, $\sigma^a$ are the Pauli matrices,   $V(\Phi^{\dagger} \Phi)=\mu^2 \Phi^{\dagger} \Phi + \lambda (\Phi^{\dagger} \Phi)^2$ ($\lambda > 0$) is the Higgs self-interaction potential and summation over repeated indices is implied. In the broken phase, $\mu^2 <0$, a vev $v/\sqrt{2}$ of the Higgs field arises classically with $v^2=-\mu^2/\lambda$. We focus on the dynamics of this system assuming that the Higgs field fluctuates slightly around its vev. In this case, we perform the standard gauge selection and we expand $\Phi$ as:
$\Phi=\left( 0, \frac{1}{\sqrt{2}}(v+  H) \right)^T$
obtaining the following equations of motion for the fields $A^{a}_{\mu}$ and $H$:
\begin{eqnarray}
(\Box &+&{ g^2 v^2 \over 4}) A^a_{\mu} - \partial_{\mu} (\partial_{\nu} A^{a, \nu}) +  {v g^2\over 2}  H A^a_{\mu} + {g^2 \over 4}  H^2 A^a_{\mu} \nonumber \\
&&+g \epsilon_{abc} [(\partial_{\mu} A^{c,\nu}) A^b_{\nu} - (\partial_{\nu} A^{b,\nu}) A^c_{\mu} - 2 A^b_{\nu}
\partial^{\nu} A^c_{\mu}] 
\nonumber \\
&&-g^2 [A^a_{\mu} A^b_{\nu} A^{b,\nu} - A^b_{\mu} A^a_{\nu} A^{b,\nu}]=0, \label{eq:eq2}\\
(\Box &+& 2 \lambda v^2)H + 3 \lambda v H^2 + \lambda H^3 \nonumber \\
&-& {g^2 \over 4} A^a_{\mu} A^{a,\mu} (v+H)=0.
\label{eq:eq3}
\end{eqnarray}
The above equations, due to the presence of a small fluctuating field $H$, suggest the use of a perturbation method with a small parameter $\epsilon$ related to the ratio of the amplitude of the Higgs field to its vev. Here, we will employ the method of multiple scales \cite{Jeffrey1982}, thus expanding the fields, variables and operators in Eqs.~(\ref{eq:eq2})-(\ref{eq:eq3}) in powers of $\epsilon$ as follows:
\begin{eqnarray}
A^a_{\mu}&=&A^a_{\mu}(0) + \epsilon A^a_{\mu}(1) + \epsilon^2 A^a_{\mu}(2) + \ldots \phantom{aaaaaaaaaaaa}\nonumber \\
H&=&H(0) + \epsilon H(1) + \epsilon^2 H(2) + \ldots \phantom{aaaaaaaaaaa}\nonumber \\
~~\frac{\partial}{\partial x^{\mu}}&=&\sum_{i=0}^{\infty} \epsilon^i \frac{\partial}{\partial x^{\mu}_i}; \qquad  x^{\mu}_i=\epsilon^i x^{\mu},
\label{eq:eq4}
\end{eqnarray}
where $j=0,1,\ldots$ in $A^a_{\mu}(j)$ and $H(j)$ denotes the order of approximation. Within this perturbative scheme, Eq.~(\ref{eq:eq3}) implies -- for reasons of self-consistency and stability -- that the fields $A^a_{\mu}$ and $H$ should be expanded around the stable minimum, $A^a_{\mu}(0)=H(0)=0$, of Eqs.~(\ref{eq:eq2})-(\ref{eq:eq3}). Generally, the dynamics of the gauge fields, without Higgs, are chaotic \cite{Savvidy1981,Wellner1992}. However, there exist configurations which admit regular solutions. For example, such a typical configuration is obtained through the color isotropic ``hedgehog" ansatz $A^a_0=0$ and $A^a_i= \delta^a_i A$ \cite{Savvidy1979,Smilga2001,Frasca2006}. In particular, such an ansatz allows the mapping of the gauge field theory to the scalar $\phi^4$ theory \cite{Frasca2006}. Here we will use a less restrictive representation of the gauge field, assuming that the non-diagonal terms are of higher-order than the diagonal terms. Furthermore, for consistency reasons implied by the structure of the equations of motion, we have to choose the temporal components of the gauge fields $A^a_0$ to be an order of magnitude larger than the other non-diagonal terms. Thus, the gauge fields can be expressed as follows:
\begin{eqnarray}
A^1_1, A^2_2, A^3_3 = A = O(\epsilon), \quad A^1_0, A^2_0, A^3_0 = O(\epsilon^2) \nonumber \\
A^1_2, A^1_3, A^2_1, A^2_3, A^3_1, A^3_2 = O(\epsilon^{\nu}), \qquad \nu \geq 3.
\label{eq:eq5}
\end{eqnarray}
The above ansatz, combined with a suitable choice for the Higgs field (which will be discussed below), leads to the decoupling of the equations of motion up to the order $O(\epsilon^3)$.

In addition, the Lorentz condition $\partial_{\mu} A^{a,\mu}=0$ is fulfilled up to the same order, $O(\epsilon^3)$. For the Higgs field we will consider two scenarios. The straightforward case $H(1) \neq 0$ leads to a set of coupled nonlinear partial differential equations (PDEs) of the NLS form for which, however, the nonlinear plane wave solutions are unstable for all the values of the relevant parameters. The details of this analysis are shown in the Appendix. In the following, we will focus on the scenario where $H(1)=0, H(2)\neq 0$: in this case, we show that nonlinear plane waves, both for the gauge and the Higgs field, are stable if the mass parameter of the Higgs field is suitably bounded.

{\it The NLS equation for the SU(2) gauge boson.}
Using the assumptions (\ref{eq:eq5}) the evolution equations (\ref{eq:eq2},\ref{eq:eq3}) for the gauge and Higgs fields, containing all contributions up to $O(\epsilon^3)$, are simplified as follows:
\begin{eqnarray}
( \Box + m_A^2 ) A &+&  {v g^2 \over 2} H A + 2 g^2 A^3 = O(\epsilon^4), \label{eq:eq6} \\
( \Box + m_H^2 ) H &+& {3\over 4}v g^2   A^2 = O(\epsilon^4),
\label{eq:eq7}
\end{eqnarray}
where
$$m_A^2={g^2 v^2 \over 4}, \quad m_H^2=2 \lambda v^2.$$
Notice that we obtain a single equation for the gauge fields due to the choice (\ref{eq:eq5}) as well as the equality of the $A^i_i$ components. Equation~(\ref{eq:eq6}) leads, to $O(\epsilon)$, to the following solution for the gauge field $A$:
\begin{equation}
A(1)=f e^{-i m_A t} + {\rm c.c.},
\label{eq:eq8}
\end{equation}
where ``c.c.'' denotes complex conjugate and the function $f=f(\vec{x}_1,t_2,...)$ is obtained from the secular terms
of Eq.~(\ref{eq:eq6}) at $O(\epsilon^2)$.
At the same order, Eq.~(\ref{eq:eq7}) determines the Higgs field:
\begin{equation}
H(1)=B\left[b \vert f \vert^2 + f^2 e^{-2 i m_A t} + (f^*)^2 e^{2 i m_A t}\right],
\label{eq:eq9}
\end{equation}
where $b= 2(m_H^2 - 4 m_A^2)/m_H^2$ and $B= -6m_A^2/bvm_H^2$. The function $f$ is determined from Eq.~(\ref{eq:eq6})
at the order $O(\epsilon^3)$, which leads to the following NLS equation:
\begin{equation}
  i \frac{\partial f}{\partial t_2} + {1 \over 2 m_A}\nabla^2_1 f  + s  \vert f \vert^2 f = 0,
\label{eq:eq10}
\end{equation}
with $s= - 2 g^2 (3 + \alpha )/ (2 m_A) $. The parameter $\alpha= (1/4) v B (b+1)$, which depends only on the ratio $q =m_H/m_A$, is given by:
\begin{equation}
\alpha(q)= -{3\over 4} \left( \frac{2}{q^2} + \frac{1}{q^2 - 4}\right).
\label{eq:eq11}
\end{equation}
The solutions of the NLS Eq.~(\ref{eq:eq10}), as well as their stability, can now be used to investigate, at the classical level, the effect of nonlinearity on the properties of the Higgs particle. In that regard, first we note that the properties of the solutions of the NLS model depend on the relative sign of the coefficients of the kinetic and nonlinear terms in Eq.~(\ref{eq:eq10}). In the case $s<0$, plane wave solutions of the NLS Eq.~(\ref{eq:eq10}) are stable (see below). Furthermore, stable localized nonlinear excitations on top of these plane waves are possible too: these include dark solitons (i.e., density dips with a phase jump across their density minima) in one-dimension (1D) \cite{Frantzeskakis2010}, vortices in two-dimensions (2D) \cite{Pismen1999} and vortex rings in three-dimensions (3D) \cite{Jones1982}. On the other hand, if $s>0$, the plane wave solutions of the NLS Eq.~(\ref{eq:eq10}) are unstable, while stable localized solutions exist only in 1D: these solutions are non-topological solitons (also known as ``bright solitons'') in the form of localized humps with vanishing conditions at infinity. Nevertheless, in higher-dimensions, all localized structures are subject to collapse -- see, e.g., \cite{Sulem1999}. From the above comments it turns out that, within this setup, there are no stable {\it and} localized classical solutions in 3D which can be interpreted -- in an obvious way -- as the elementary Higgs particle.

In the following, we focus on the plane wave solutions of Eq.~(\ref{eq:eq10}) and their stability. As stated above, these solutions are particularly important for the effective description of the spontaneously broken vacuum and excitations on top of it, or in their use as a basis for the quantization of the theory.

{\it Stability constraints and the Higgs mass}.

The NLS Eq.~(\ref{eq:eq10}) possesses exact analytical plane wave solutions of the form:
\begin{equation}
f(\vec{x}_1,t_2)=f_0 \exp[-i(\omega t_2- \vec{k}\cdot\vec{x}_1)],
\label{eq:eq12}
\end{equation}
where $f_0$ is the amplitude of the plane wave for the gauge field, while its frequency
$\omega$ and the wave vector $\vec{k}$ satisfy the following dispersion relation:
\begin{equation}
\omega(k)={\vec{k}^2 \over 2 m_A} - s|f_0|^2.
\label{eq:eq13}
\end{equation}
The solutions (\ref{eq:eq12}) resemble the plane wave solutions usually employed for the quantization of the SU(2)-Higgs theory, but they feature a modified dispersion relation, cf. Eq.~(\ref{eq:eq13}), involving an amplitude-dependent mass correction. This correction originates from the self-interaction, as well as the interaction between the gauge and the Higgs fields, and has, in principle, a similar effect as the quantum self-energy correction. Similar solutions have been found in Ref.~\cite{Savvidy1979} for the pure SU(2) model in the vanishing ``Poynting'' vector frame of reference.

The stability of the nonlinear plane waves (\ref{eq:eq12}) can be studied by considering small-amplitude perturbations which, for a static background ($k = 0$), assume the form:
\begin{equation}
\delta f=(u+i v) \exp[-i((\Omega+\omega) t_2 -\vec{Q} \vec{x}_1)]
\label{eq:eq15}
\end{equation}
In Eq.~(\ref{eq:eq15}), $u$ and $v$ denote the (real) amplitudes, while  $\Omega$, $\vec{Q}$ the frequency and wave vector of the perturbation, respectively \cite{comment}; then, it is straightforward to find from Eq.~(\ref{eq:eq10}) that $\Omega$ and $\vec{Q}$ satisfy the dispersion relation:
\begin{equation}
\Omega^2 = \frac{|\vec{Q}|^2}{2 m_A} \left(-2 s |f_0|^2 + \frac{ |\vec{Q}|^2}{2 m_A}\right).
\label{eq:eq16}
\end{equation}
Hence, the solution (\ref{eq:eq12}) is stable, i.e., $\Omega$ is real for all
$|\vec{Q}|$, only if $s<0$, i.e., if the parameter $\alpha$ obeys:
\begin{equation}
3+\alpha(q) >0.
\label{eq:eq17}
\end{equation}
The above inequality determines the permitted regions for the ratio $q$ -- thus restricting $m_H$ for given $m_A$ -- so that plane waves of the gauge field are stable. 
Thus, although the Born approximation of the usual perturbation theory  
(which neglects all nonlinear terms for small coupling) does not imply any information concerning the properties of the Higgs field, in the framework of multi-scale analysis we obtain restrictions for the Higgs mass originating from stability conditions.

Interestingly enough, we will now show that these restrictions do not exclude the region, which turns out to be the most probable for the Higgs mass based on the recent experimental observations. Particularly, 
utilizing the Standard Model value: $m_A=80$~GeV we find that $m_H$ fulfils the condition:
\begin{equation}
56~{\rm GeV} <m_H < 160~{\rm GeV} \quad {\rm or} \quad m_H > 165~{\rm GeV}.
\label{eq:eq18}
\end{equation}
The region of very low Higgs masses ($m_H < 114.5~{\rm GeV}$) is practically excluded from existing experimental data (LEP II \cite{LEP}). In addition, the zone $m_H \in [158,173]~{\rm GeV}$ has been excluded from the Tevatron analysis \cite{Tevatron}. Latest results from ATLAS and CMS experiments at LHC-CERN \cite{LHC}, when combined, practically exclude the region $[145,466]~{\rm GeV}$. Thus, the most probable scenario for a light Higgs field, compatible with existing experimental data and the above analysis, is $114.5~{\rm GeV} < m_H < 145~{\rm GeV}$. According to the Standard Model, in this range of the Higgs mass parameter, the decay of Higgs into a pair of opposite charged gauge bosons starts to increase rapidly \cite{Djouadi1998}, dominating for $m_H > 140~{\rm GeV}$ over the other channels. This partly justifies the use of the simplified model considered here, which takes into account the interaction of the Higgs field with the gauge bosons while neglecting all fermion couplings.

{\it Concluding remarks}.
We have employed a multiple scale perturbative scheme to analyze the classical SU(2)-Higgs dynamics, taking into account 
the nonlinear terms in the corresponding equations of motion. This approach enables a consistent treatment of the secular terms occurring at higher orders of the classical perturbation theory. Using a suitable representation for the gauge field, the equations of motion of the complete SU(2)-Higgs model simplify significantly and are reduced, depending on the magnitude of the Higgs field amplitude, either to a single NLS equation (weak Higgs field, $\propto \epsilon A$) or to a set of two coupled NLS equations (strong Higgs field, $\propto A$ -- see Appendix). These equations admit nonlinear plane wave solutions, with an amplitude-dependent dispersion relation, in contrast to the ones usually used for the quantization of the model. The relevance of nonlinear wave solutions for the field quantization has been recently discussed in Ref.~\cite{Himpsel2011}. These plane wave
solutions are stable for a range of Higgs mass  values and only in the case of a weak Higgs field. On the other hand, for a strong Higgs field the nonlinear plane waves are unstable for all parameter values. Thus, the classical stability analysis of the nonlinear solutions of the reduced SU(2)-Higgs evolution equations provides bounds in the characteristics 
(amplitude, mass) of the Higgs field, which could be of relevance for the running experimental studies. Finally, one should notice that the dynamics of the coupled NLS equations occurring in the case of a strong Higgs field (see Appendix) require a more extensive analysis in order to explore also other solutions which could be relevant for the dynamical description of the SU(2)-Higgs model. This is an interesting direction for future studies.

\vspace{+12pt}
\noindent {\bf Acknowledgments.} We thank P. G. Kevrekidis for extensive, helpful and enlightening discussions.

\appendix
\section{The case of strong Higgs field amplitude}

Here we study the case where the expansion of the Higgs field in powers of $\epsilon$ 
starts from the first order, i.e., the case when the ratio of the Higgs to the gauge boson field amplitude is larger. 
This way, we consider the following asymptotic expansion of $H$:
\begin{equation}
H = \epsilon H(1) + \epsilon^2 H(2) + \ldots, 
\label{eq:eq19}
\end{equation}
and keep all other expressions in Eq.~(\ref{eq:eq5}) the same as in the preceding analysis. Then Eqs.~(\ref{eq:eq6})-(\ref{eq:eq7}) for the gauge  and Higgs fields are modified and their form, containing all contributions up to $O(\epsilon^3)$, is:
\begin{equation}
( \Box + m_A^2 ) A + {{g^2 v} \over 2}H A + {g^2 \over 4} H^2 A + 2 g^2 A^3 = O(\epsilon^4),
\label{eq:eq20}
\end{equation}
\begin{equation}
( \Box + m_H^2 ) H + \lambda H^2 (H + 3 v) + {3 g^2 A^2 \over 4} ( v + H ) = O(\epsilon^4).
\label{eq:eq21}
\end{equation}
The perturbative treatment of Eqs.~(\ref{eq:eq20}) and (\ref{eq:eq21}) at each order of 
$\epsilon$ leads to the the following form for the fields $A(1)$ and $H(1)$:
\begin{eqnarray}
A(1)&=&f(\vec{x}_1,t_2,\ldots) e^{-i m_A t}+{\rm c.c.}, \nonumber \\
H(1)&=&h(\vec{x}_1,t_2,\ldots) e^{-i m_H t}+{\rm c.c.}.
\label{eq:eq22}
\end{eqnarray}
The functions $f$ and $h$ satisfy two coupled NLS equations of the following form:
\begin{eqnarray}
i\partial_{t_2} f &+& {1 \over 2 m_A} \nabla^2_1 f - [g_{11}|f|^2 + g_{12}|h|^2 ]f=0, \nonumber \\
i\partial_{t_2} h &+& {1 \over 2 m_H} \nabla^2_1 h - [g_{21}|f|^2 + g_{22}|h|^2 ]h=0,
\label{eq:eq23}
\end{eqnarray}
where the coefficients $g_{ij}$ ($i,j \in \{1,2\}$) are given by
\begin{eqnarray}
g_{11}&=&{g^2 \over m_A} (3 + \alpha(q)),\quad  \quad g_{12}={g^2 \over 4m_A}
\theta(q), \nonumber \\
g_{21}&=&{3 g^2 \over 4 m_H}  
\theta(q), \quad \quad \quad  g_{22}=-{3 g^2 \over 4 m_H} q^2,
\label{eq:eq24}
\end{eqnarray}
the parameter $\alpha(q)$ is given by Eq.~(\ref{eq:eq11}), and $\theta(q)=-3 + q^2/(q^2 -4)$.

The system of equations (\ref{eq:eq23}) admits plane wave solutions of the form:
\begin{eqnarray}
f(\vec{x}_1,t_2)&=&f_0 \exp[-i(\omega_f t_2 -\vec{k}_f \vec{x}_1)], 
\nonumber \\
h(\vec{x}_1,t_2)&=&h_0  \exp[-i(\omega_h t_2 -\vec{k}_h \vec{x}_1)],
\label{eq:eq25}
\end{eqnarray}
where the frequencies $\omega_f ,\omega_h$  satisfy the following dispersion relations:
\begin{eqnarray}
\omega_f &=& {\vec{k}_f^2 \over 2 m_A} + g_{11}\vert f_0 \vert^2 + g_{12} \vert h_0 \vert^2, \nonumber \\
\omega_h &=& {\vec{k}_h^2 \over 2 m_H} + g_{21}\vert f_0 \vert^2 + g_{22} \vert h_0 \vert^2.
\label{eq:eq26}
\end{eqnarray}
In order to investigate the stability of the solutions (\ref{eq:eq25}) on top of a static background, i.e., $k_f=k_h=0$, we introduce small perturbations of the form:
\begin{eqnarray}
\delta f&=& (u_1+i v_1) \exp[-i((\Omega + \omega_f) t_2 -\vec{Q} \vec{x}_1)],  
\nonumber \\
\delta h&=&(u_2+i v_2) \exp[-i((\Omega + \omega_h) t_2 -\vec{Q} \vec{x}_1)],
\label{eq:eq27}
\end{eqnarray}
where the perturbation amplitudes $u_i$, $v_i$ ($i=1,2$) are assumed to be real, while  
$\Omega$, $\vec{Q}$ 
denote the energy and the momentum of the perturbations, respectively. Applying again the method used above for the weaker Higgs field, we require that the small variations (\ref{eq:eq27}) do not diverge with time. This is satisfied if the roots of the bi-quadratic equation:
\begin{equation}
\Omega^4-b\Omega^2+c=0,
\label{eq:eq29}
\end{equation}
are real. The coefficients $b$ and $c$ read:
\begin{eqnarray}
b&=&p_H b_H+p_A b_A,  \nonumber \\
c&=&p_H p_A b_H b_A -4p_H p_A \vert f_0 \vert^2 \vert h_0 \vert^2 g_{12}g_{21},
\label{eq:eq30}
\end{eqnarray}
where 
$p_H= \vec{Q}^2/2m_H$, $p_A= \vec{Q}^2/2m_A$, $b_H=p_H+2g_{22} \vert h_0 \vert^2$, and $b_A=p_A+2g_{11} \vert f_0 \vert^2$. 
Real roots of (\ref{eq:eq29}) imply $b>0$ and $c>0$, leading to the following stability conditions:
\begin{equation}
g_{11}g_{22}-g_{12}g_{21} > 0, \quad  g_{22} > 0, \quad g_{11} > 0. 
\label{eq:eq31}
\end{equation}
For the coefficients $g_{ij}$ [cf. Eq.(\ref{eq:eq24})] there are no values of $q$ satisfying the first two inequalities. Thus, according to this analysis, for a stronger Higgs field, plane wave solutions are unstable.

\end{document}